\documentclass[prb,twocolumn,preprintnumbers,amsmath,amssymb]{revtex4}
\usepackage{CJK}
\pdfoutput=1
\usepackage{graphicx}% Include figure files
\usepackage{dcolumn}% Align table columns on decimal point
\usepackage{bm}% bold math
\usepackage{upgreek}
\usepackage{amsmath}
\usepackage{url}
\usepackage{hyperref}
\usepackage{multirow}

\newcommand{\beq}{\begin{equation}}
\newcommand{\eeq}{\end{equation}}
\newcommand{\beqa}{\begin{eqnarray}}
\newcommand{\eeqa}{\end{eqnarray}}

\begin{document}
\title{Impact of ex-situ rapid thermal annealing on the magneto-optical properties and the oscillator strength of In(Ga)As quantum dots}% Force line breaks with \\

\author{T. Braun$^1$}
\author{S. Betzold$^1$}
\author{N. Lundt$^1$}
\author{M. Kamp$^1$}
\author{S. H\"ofling$^{1,2}$}
\author{C. Schneider$^1$}

\affiliation{$^1$Technische Physik and Wilhelm Conrad R\"ontgen Research Center for Complex Material Systems, Physikalisches Institut,
Universit\"at W\"urzburg, Am Hubland, D-97074 W\"urzburg, Germany $^2$ SUPA, School of Physics and Astronomy, University of St. Andrews, St. Andrews KY 16 9SS, United Kingdom}

\date{\today}

\begin{abstract}
We discuss the influence of a rapid thermal annealing step on the magneto-optical emission properties of In(Ga)As/GaAs quantum dots. We map out a strong influence of the growth- and anneling parameters on the quantum excitons' effective Land\'e g-factors and in particular on their diamagnetic coefficients, which we directly correlate with the modification of the emitters shape and material composition. In addition, we study the excitons' spontaneous emission lifetime as a function of the annealing temperature and the dot height, and observe a strong increase of the emission rate with the quantum dot volume. The corresponding increase in oscillator strenth yields fully consistent results with the analysis of the diamagenic behavior. In particular, we demonstrate that a rapid thermal annealing step of 850$^\circ$ C can be employed to increase the oscillator strength of as-grown InAs/GaAs QDs by more than a factor of $2$.
\end{abstract}
\maketitle

\section{Introduction}
Tremendous progress has been achieved the last years in understanding and exploiting the electronic as well as the photonic characteristics of semiconductor quantum dots (QDs), resulting in remarkable achievements such as the observation of triggered single photon emission\cite{Michler2000,Santori2001,Moreau2001}, generation of highly indistinguishable\cite{Santori2002, He2013} and entangled\cite{Akopian2006} photon pairs as well as reaching the strong coupling regime of a single solid state quantum emitter and the localized field of a cavity mode\cite{Reithmaier2004, Yoshie2004, Peter2005}. The QD's oscillator strength (OS) is a dimensionless parameter describing the strength of the coupling of a transition from an excited to the ground state of a quantum emitter to the surrounding light field and is therefore a crucial factor in any kind of cavity quantumelectrodynamics (cQED) experiment. A larger OS leads to a stronger light matter interaction (light matter coupling constant $g\propto \sqrt{f}$) which is, e.g. crucial to reach the strong coupling regime where a coherent exchange of light and matter is possible. In the limit of weak confinement of an exciton in a QD, the OS is associated to the QD's size. Thus, morphologically large QDs in the III/V semiconductor system such as low strain InGaAs QDs ($f$ up to 50)\cite{Reithmaier2004} and natural GaAs islands ($f\approx 100$)\cite{Peter2005} have been identified to particularly promote strong coupling conditions. As these QDs evolve just at the layer-to-island transition of the Stranski-Kranstanov growth mode, or form as natural islands driven by surface migration, routinely reproducing and engineering QDs with a high OS is very difficult.\\
Several groups have shown that rapid thermal annealing (RTA) of InAs QDs is a useful technique to take influence on the emission properties after the growing process. For example, RTA allows for tailoring the QD size, the indium composition and with that the emission energy as well as the linewidth of the quantum dot ensemble\cite{Kosogov1996,leon1996effects,Fafard1999,Nishi1999,Hsu2000}. Furthermore it is possible to adjust the dephasing time of QD excitons\cite{Langbein2004}. Recently, it also has been indicated that post growth, rapid thermal annealing of InAs QDs can also be exploited to significantly enhance the oscillator strength of QDs\cite{Loo2010}. While these results are very promising, a thorough investigation of the effect of rapid thermal annealing on the intrinsic excitonic properties is still elusive. Here, in a systematic analysis we demonstrate the influence of a rapid thermal annealing (RTA) process on the QD properties via magnetic field dependent micro-photoluminescence (�PL) and time resolved spectroscopy. We show an increase of the mean lateral expansion of the exciton wavefunction in InAs/GaAs QDs by about 70\% and more than a doubling of the OS under appropriate growth and annealing conditions. Furthermore, we demonstrate that measuring the diamagnetic response of a QD exciton is a viable tool to directly assess the emitter's oscillator strength  as it yields reliable and consistent results on the single emitter scale.

\section{Theory}
\label{Theory}

To describe a QD in a magnetic field on a basic level, we consider the exciton energy $E_{0}$ without magnetic field, the Zeeman interaction between electrons and holes and the diamagnetic shift energy which is induced by the external magnetic field. The total energy of the system in a magnetic field in Faraday configuration (${\bf B}=(0,0,B_z)$) is then given by:

\begin{align}
E_{X}(B_z)&=E_{0}+E_{Zeeman}+E_{Dia}\nonumber \\ 
&=E_{0}\pm\frac{1}{2}\mu_{B}g_{X}B_z+\chi B_z^2~,
\label{eq:Exzitonenergie_Magnetfeld}
\end{align}

with the Bohr magneton $\mu_B$ and an effective excitonic Land\'e-factor $g_X$ in the Zeeman-term. The diamagnetic coefficient $\chi$ is defined by\cite{Babinski2006}:

\beq
\chi=\frac{e^2}{8}\left(\frac{\left<\rho_e^2\right>}{m_{e,eff}}+\frac{\left<\rho_h^2\right>}{m_{h,eff}}\right)~.
\label{eq:diamagnetischer_Koeffizient}
\eeq

Hereby $\left<\rho^2_i\right>=\left<x_i^2+y_i^2\right>$ gives the mean quadratic expansion of the wave function of the carriers.

The dimensionless oscillator strength $f$ describes the coupling strength between the light field and the transition from an excited to the ground state of the emitter. It indicates the absorption and recombination probability in bulk material and depends primarily on the overlap of the wave functions of the electron and the hole. $f$ can be correlated with the radiative decay rate $\Gamma(\omega)$ via\cite{Hours2005}:

\beq
f(\omega)=\frac{6\pi m_0 \epsilon_0 c^3}{n(\omega)e^2\omega^2}\Gamma(\omega)~,
\label{eq:Oszillatorst�rke_lang}
\eeq

where $n(\omega)$ is material's the refractive index. For QDs which are larger than the excitonic Bohr radius $a_B$ (this is the so-called limit of weak confinement), $f$ can be approximated by\cite{Stobbe2010}:

\beq
f(\omega,x)=\frac{2E_P(x)}{\hbar\omega}\left(\frac{L}{a_B}\right)^2~.
\label{eq:Oszillatorst�rke}
\eeq

Here $E_P(x)$ is the Kane-energy\cite{Kane1957} and $L$ the expansion of the center-of-mass wave function in the plane perpendicular to the growth direction of the QD.  

\section{Sample design and experimental methods}
\subsection{Sample preparation}
\begin{figure}[t]
\begin{center}
\includegraphics[width=0.48\textwidth]{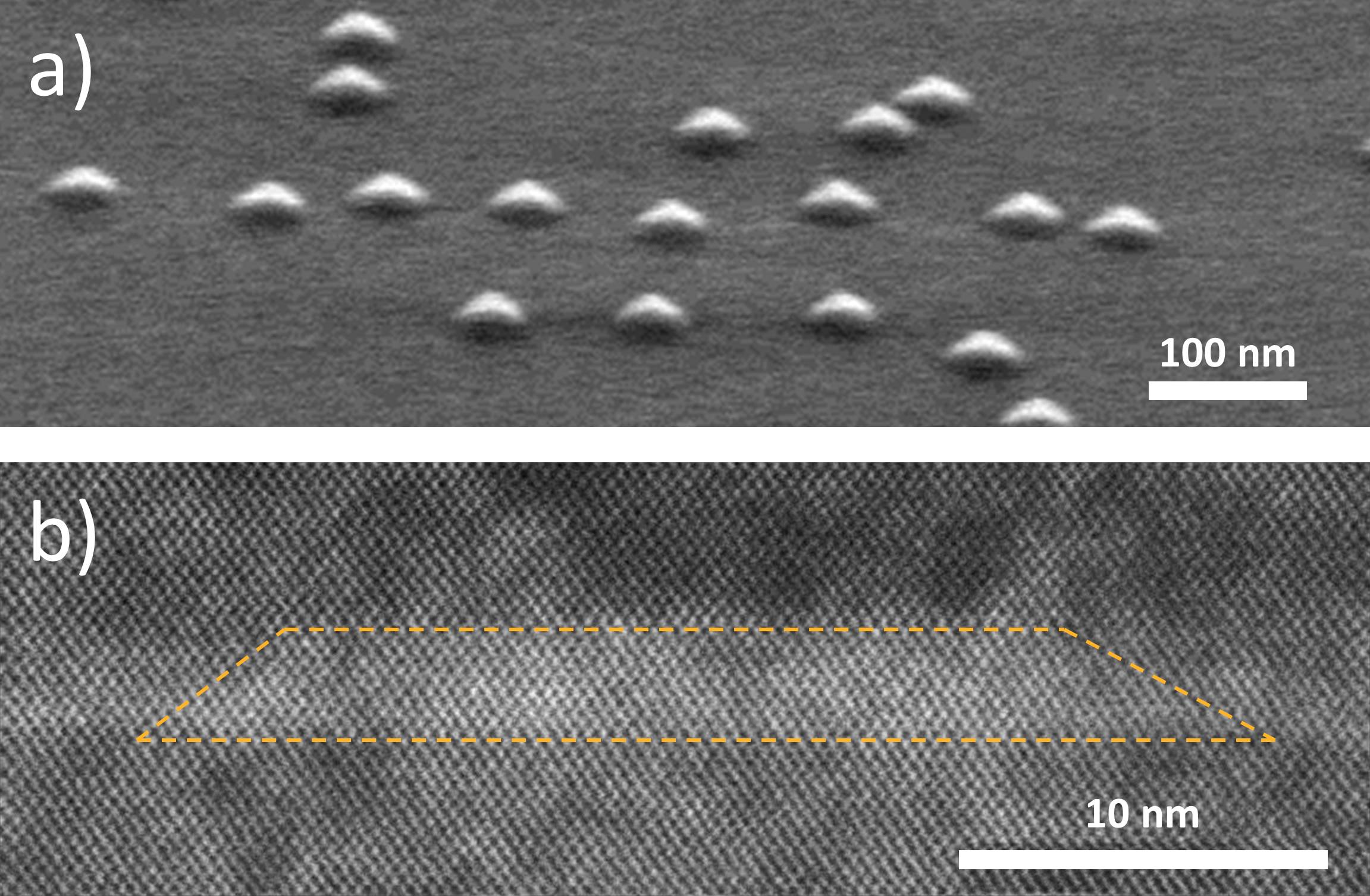}
\caption{(a) SEM image of surface QDs before the capping process. (b) STEM image of a QD after partial capping ($d=3~nm$) and annealing and overgrowth by 50 nm of GaAs.}
\label{fig:M4045}
\end{center}
\end{figure}

The QDs under consideration were grown via molecular beam epitaxy (MBE) and treated by the partial capping and annealing (PCA) technique. The InAs-QDs deposited on a GaAs buffer layer were overgrown by a GaAs cap layer and were then in-situ annealed for one minute at a nominal substrate temperature of 560$^\circ$~C. The PCA process leads to an enhanced interdifusion of the Ga- and In-atoms and flushes away the uncovered QD tip\cite{Garcia1997}. The QD height and hence the emission energy hereby strongly depend on the cap thickness. After the in-situ annealing process the crystal growth was completed by a 100 nm thick GaAs cap to circumvent nonradiative surface recombination. The emission frequency of the QDs can additionally be tuned by an ex-site rapid thermal annealing (RTA) step. Thereby the samples were heated up to temperatures in the range of  750$^\circ$~C to 850$^\circ$~C  for 5 minutes. Thus the interdiffusion process is boosted again and the further mixing of the atoms leads to an increase in the QD volume. In addition a blue shift of the emission takes place due to the reduction of the In content in the QD\cite{leon1996effects,Fafard1999,Hsu2000}. For a detailed analysis of the influence of the annealing process on the OS we have grown three samples with a varying cap thickness $d$ from 2~nm to 4~nm. Each specimen was cleaved into four pieces, and three of these pieces were treated via RTA at 750$^\circ$~C, 800$^\circ$~C and 850$^\circ$~C. For the fourth piece, the annealing step was omitted and the sample served as a reference. Fig. \ref{fig:M4045} shows a scanning electron microscope (SEM) image of the surface QDs prior to the capping (a) and a scanning transmission electron microscope (STEM) image of a QD after the growth of the full structure has been accomplished (b). The thickness of the partial cap was chosen to $d=3~nm$ in this particular sample.  While the surface QDs exhibit the typical pyramidal shape, the top side of the overgrown QD is significantly flattened and QD height decreased down to the thickness of the capping layer.

\subsection{Experimental methods}
Magneto-optical studies were performed at a temperature of 5 K using a high resolution �PL setup. The  magnet cryostat provides magnetic flux densities up to 5~T in Faraday configuration. The samples were excited with a Ti:Sapphire laser tuned to 780~nm which can be operated in cw- as well as in pulsed-mode with a repetition rate of 82~MHz. The emitted photons were collected by a 50x microscope objective ($NA= 0.42$), dispersed by a grating with 1200~lines/mm and detected with a LN$_2$ cooled Si-CCD. The magneto-optical measurements were performed in cw-mode. To investigate the Zeeman-splitting, a quarter wave plate was mounted in addition to a linear polarizer. Adjusting the quarter wave plate accurately in an angle of $\pm 45^\circ$ with respect to linear polarizer we can extract the circular polarized part of the QD emission.\\
The samples were also investigated under pulsed excitation in order to measure the lifetime of the QDs. The emission signal of the quantum dot ensemble was filtered by the monochromator and then focused on a fiber-coupled avalanche photo diode (APD) with a temporal resolution of $\sim 40$~ps. 

\section{Experimental results}
\begin{figure}[t]
\begin{center}
\includegraphics[width=0.48\textwidth]{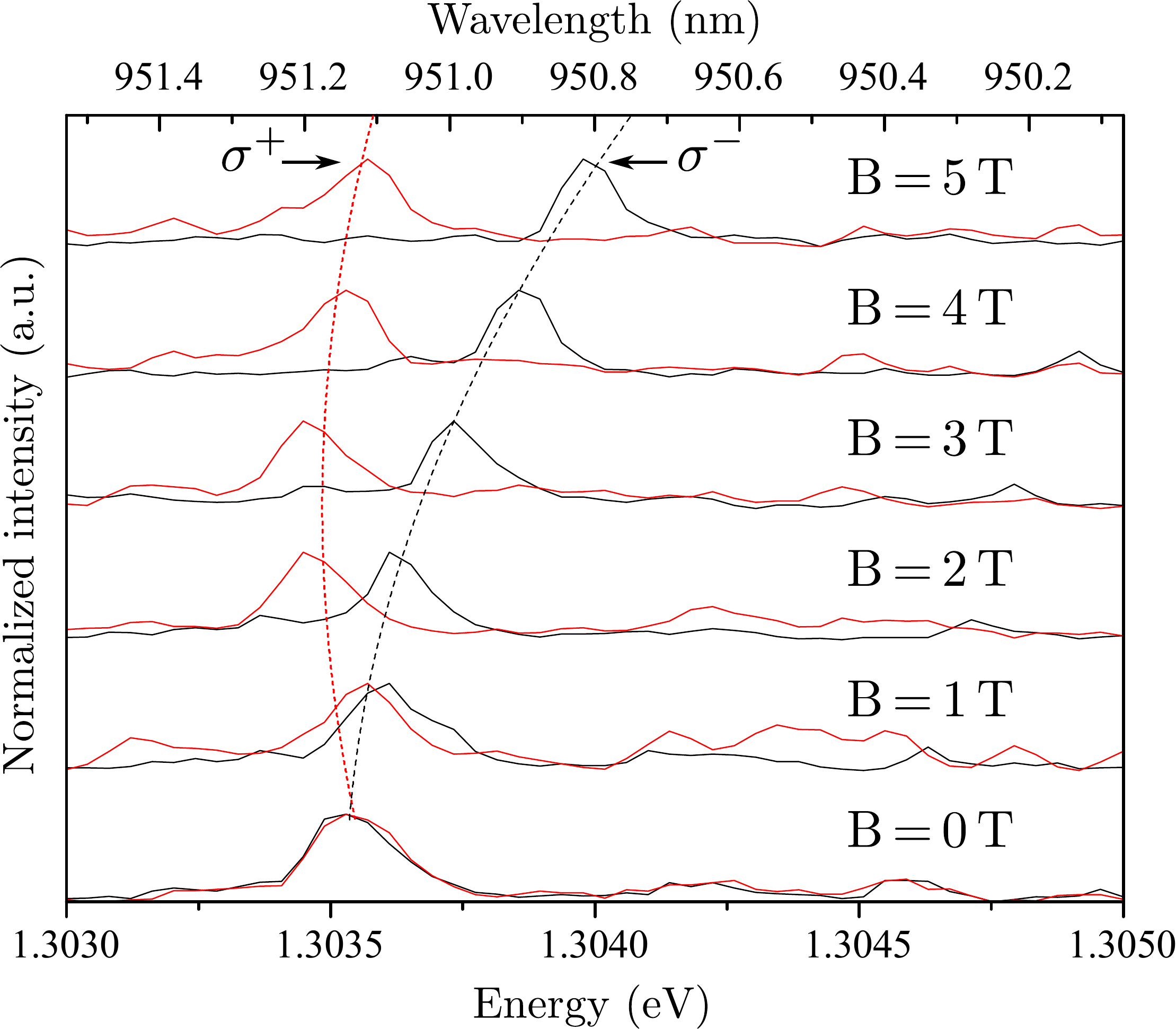}
\caption{Plot of the normalized intensity of the right- and left-circular polarized part of the emission of a QD for different magnetic fields (Reference sample with a cap thickness of $d=3~nm$). An increasing Zeeman splitting between $\sigma^+$ and $\sigma^-$ is observable as well as a superimposed diamagnetic shift. }
\label{fig:Zeeman_splitting}
\end{center}
\end{figure}

Figure \ref{fig:Zeeman_splitting} shows a typical emission spectrum of a QD at different magnetic fields. The investigated sample was the reference sample with a cap thickness of 3~nm. The intensity was normalized for each spectrum. Since the typical QD fine structure splitting (FSS) is on the order of  10~$\mu$eV for these QDs and the resolution of our setup is about 35~$\mu$eV no notable splitting was detected (without performing more sophisticated measurements). However, one clearly sees the influence of the magnetic field (cf. equation \ref{eq:Exzitonenergie_Magnetfeld}) on the QD emission energy. The Zeeman splitting between the right ($\sigma^+$)- and left ($\sigma^-$)-circular polarized photons increases linearly with the magnetic field and the diamagnetic shift superimposes the split doubled.

\subsection{Zeeman splitting and Land\'e g-factor}
\label{zeeman}

\begin{figure}[t]
\begin{center}
\includegraphics[width=0.48\textwidth]{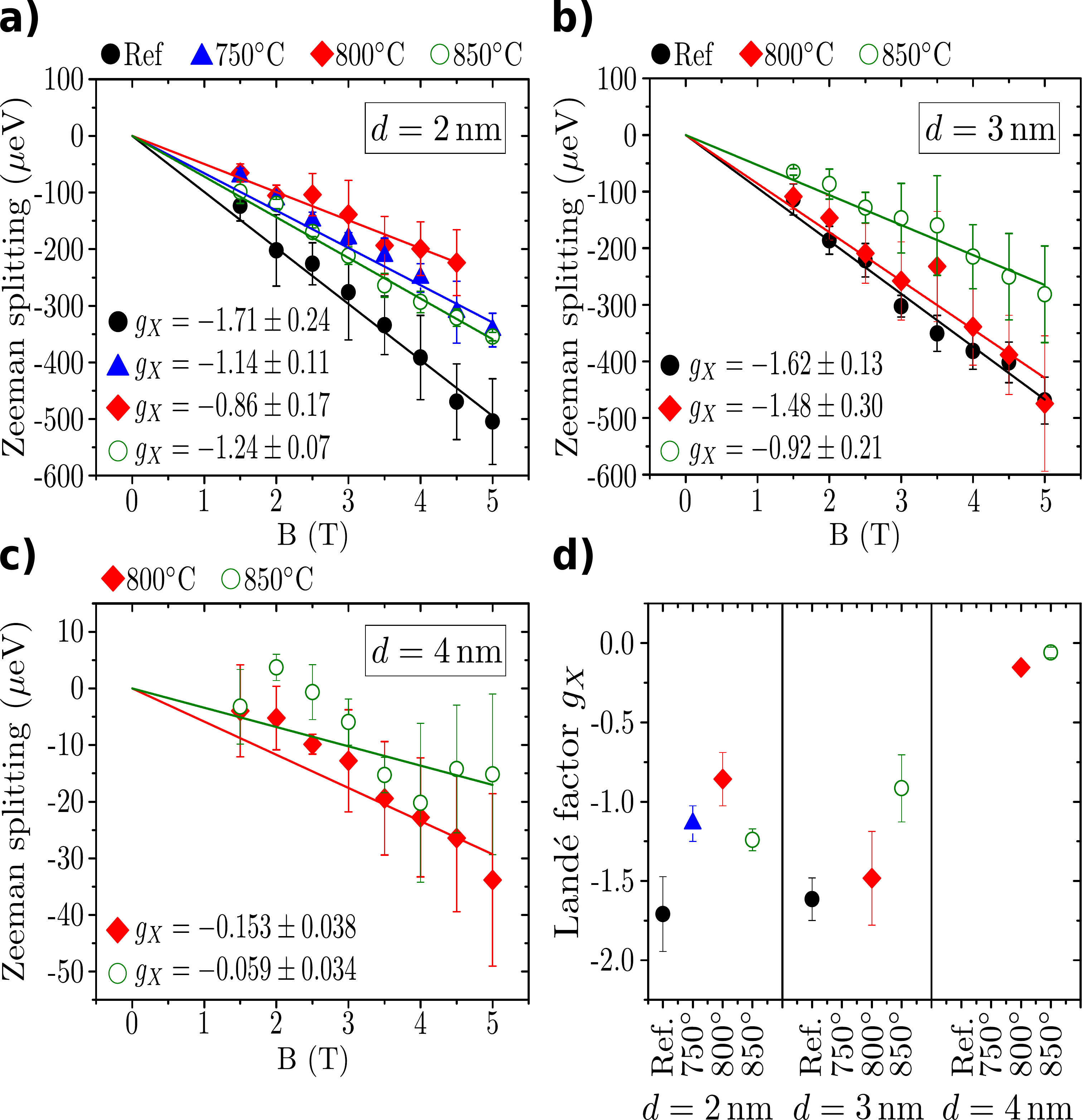}
\caption{Zeeman splitting versus the magnetic field for cap thicknesses $d$ of 2~nm a), 3~nm b) and 4~nm c) as well as the determined effective g-factors in dependence of the ex-situ annealing temperature d). The decreasing of the absolute $g_X$-value for higher annealing temperatures implies an increasing lateral expansion of the QDs and a modification of the composition and a strain variation in the QDs.}
\label{fig:zee_alle}
\end{center}
\end{figure}

In order to study the evolution of the Zeeman splitting and the effective Land\'e g-factors of the QDs in greater detail, we plot the energetic gap between the $\sigma^+$ and the $\sigma^-$ emission, which is defined as $E_{Zeeman}=E(\sigma^+) - E(\sigma^-)$, as a function of the magnetic field. The result for each sample is depicted in fig. \ref{fig:zee_alle} a)-c). Up to 9 different QDs per sample were investigated and the resulting mean value was plotted. Spectra were taken every 0.5~T up to a magnetic flux density of 5~T. We assume that the emission is completely circular polarized for $T \geq 1.5~T$\cite{Bayer2002}.

For each cap thickness a decrease of the Zeeman splitting and accordingly the g-factors $|g_X|$ is observable for increasing annealing temperatures. According to equation \ref{eq:Exzitonenergie_Magnetfeld} the effective g-factor can be obtained by a linear fit. The absolute g-value $|g_X|$ of the QDs of the sample with $d=2~nm$ drops from $1.71 \pm 0.24$ for the reference sample to $0.86 \pm 0.17$ at a temperature of 800$^\circ$~C (cf. fig. \ref{fig:zee_alle} a)). We attribute this effect to the interdiffusion of the Ga- with the In-atoms which leads to a lateral expansion of the QDs. According to Bayer et al.\cite{Bayer1995} and Kotlyar et al.\cite{Kotlyar2001} $g_X$ should be proportional to $1/D^2$ ($D$ is the QD diameter). This effect is caused by spin-orbit interactions which increase with decreasing QD size due to the uncertainty relation. Furthermore we would like to note that also the composition in the QD significantly modifies both with the annealing temperature and the dot size, and thermal annealing additionally leads to a strain variation in the QD which results in a reduction of the absolute $g_X$-value\cite{Nakaoka2005}. This is consistently observed in fig. \ref{fig:zee_alle}d, where we depict all obtained g-factors for varying cap thicknesses and annealing temperatures. The highest absolute $g_X$-value is measured for the reference sample with $d=2~nm$ with $|g_X|=1.71\pm 0.24$ and the lowest for the sample with a 4~nm cap annealed at 850$^\circ$~C with $|g_X|=0.059\pm 0.034$. We note that the emission band of the sample with a cap thickness of 4~nm was outside of the detection window of our Si-CCD camera (both for the reference sample and the sample piece that was annealed at 750$^\circ$~C).  

\subsection{Diamagnetic coefficient $\chi$}
\begin{figure}[t]
\begin{center}
\includegraphics[width=0.48\textwidth]{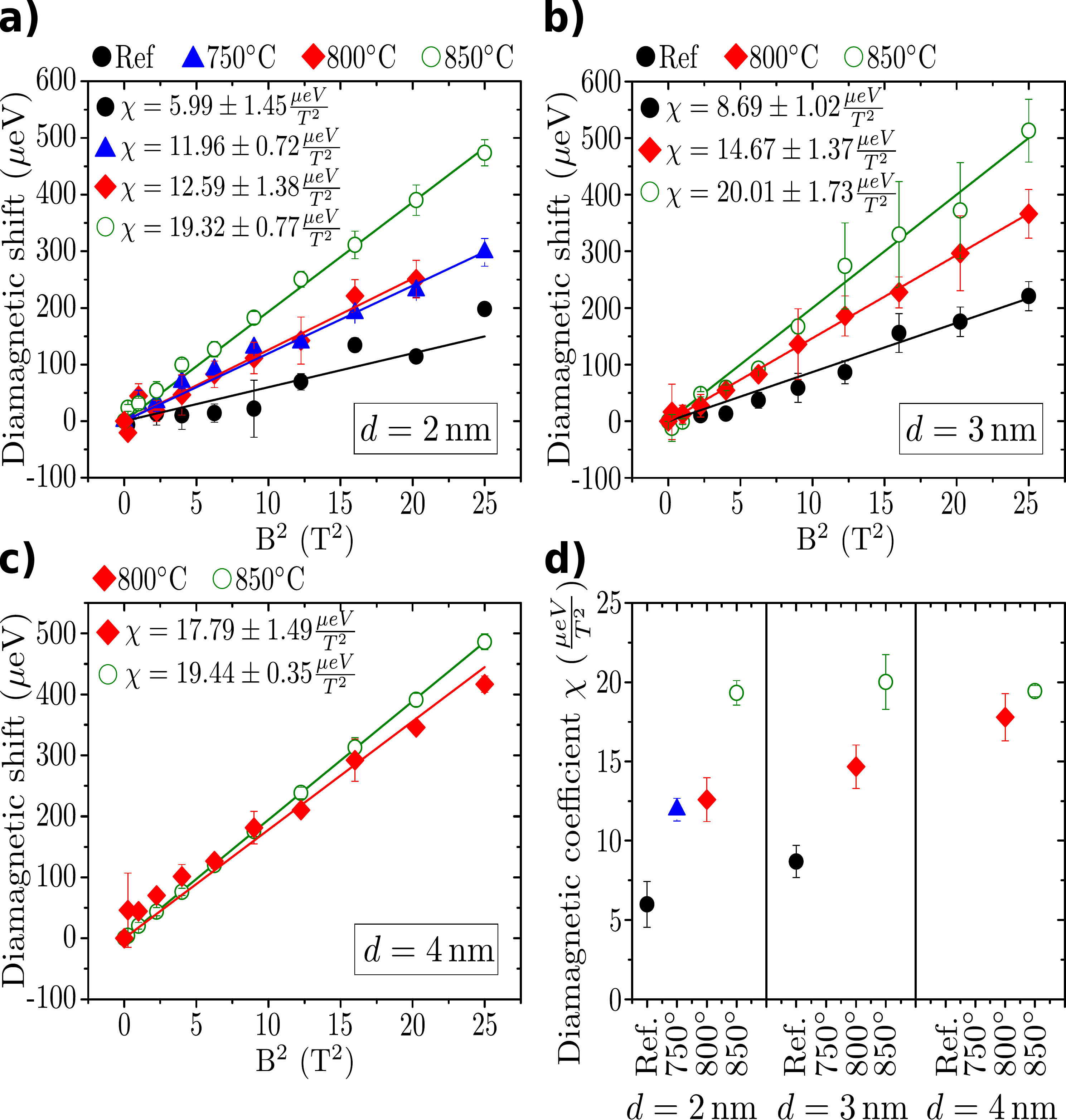}
\caption{Diamagnetic shift versus square the magnetic field for cap thicknesses $d$ of 2~nm a), 3~nm b) and 4~nm c) as well as the determined diamagnetic coefficient $\chi$ as a function of the ex-situ annealing temperature d). For increasing cap thicknesses and higher annealing temperatures a discinct increase of $\chi$ is observable.}
\label{fig:dia_alle}
\end{center}
\end{figure}

Next, we study the diamagentic behavior of the emitters as a function of the annealing conditions.  We plot the magnitude of the diamagnetic shift (which can be obtained via the energy mean value of the $\sigma^+$ and $\sigma^-$ emission lines) against the square of the magnetic field. Then a linear fit is applied to the data to extract the diamagnetic coefficient $\chi$. The results are shown in fig. \ref{fig:dia_alle}.

For all three chosen cap thicknesses we observe a distinct increase of the diamagnetic coefficient for higher ex-situ annealing temperatures. In addition, an increasing $\chi$ for the same temperatures and higher cap thicknesses can be noted (for $d=3~nm$). It is reasonable to assume that thicker caps and higher temperatures lead to an increase of the lateral expansion of the QDs since $\chi$ is proportional to the expansion of the wave functions according to equation \ref{eq:diamagnetischer_Koeffizient}.
With equation \ref{eq:Oszillatorst�rke} and the approximation of $L\approx 2 \sqrt{\left< \rho^2_{e,h} \right>}~$\cite{Reitzenstein2009} we can estimate the mean lateral expansion of the wavefunction and can calculate the OS $f$ (cf. equation \ref{eq:diamagnetischer_Koeffizient}):

\beq
f(\omega,x)=64\frac{E_P(x)}{\hbar\omega} \frac{m_{e,eff}m_{h,eff}}{e^2 a_B^2( m_{e,eff}+m_{h,eff})}
\cdot \chi~.
\label{eq:OS_dia}
\eeq

The determined values for $f$ will be discussed later in section \ref{os} and are plotted as black squares in fig. \ref{fig:Oszillatorstarke}. It should be noted that L is not equivalent to the physical expansion of our quantum dots as seen in the TEM image in figure \ref{fig:M4045}. The determined L value (8.1~nm) and the average diameter of the quantum dots taken from TEM analysis (25.8~nm) deviate significantly. In order to explain the difference we performed six-band-\textbf{\textit{k}$\cdot$\textbf{\textit p}} model based calculations of the wave functions, using the software Nextnano \cite{Trellakis2006}. For this, we used geometric parameters taken from TEM analyis. The resulting mean diameters of the wave functions are 9.8~nm for the electron wave function $\left< \rho^2_e \right>$ and 12.4~nm for the hole wave function $\left< \rho^2_h \right>$. Therefore, the significant deviation between L and D are not in contradiction.

\subsection{Time resolved spectroscopy}
\begin{figure}[t]
\begin{center}
\includegraphics[width=0.48\textwidth]{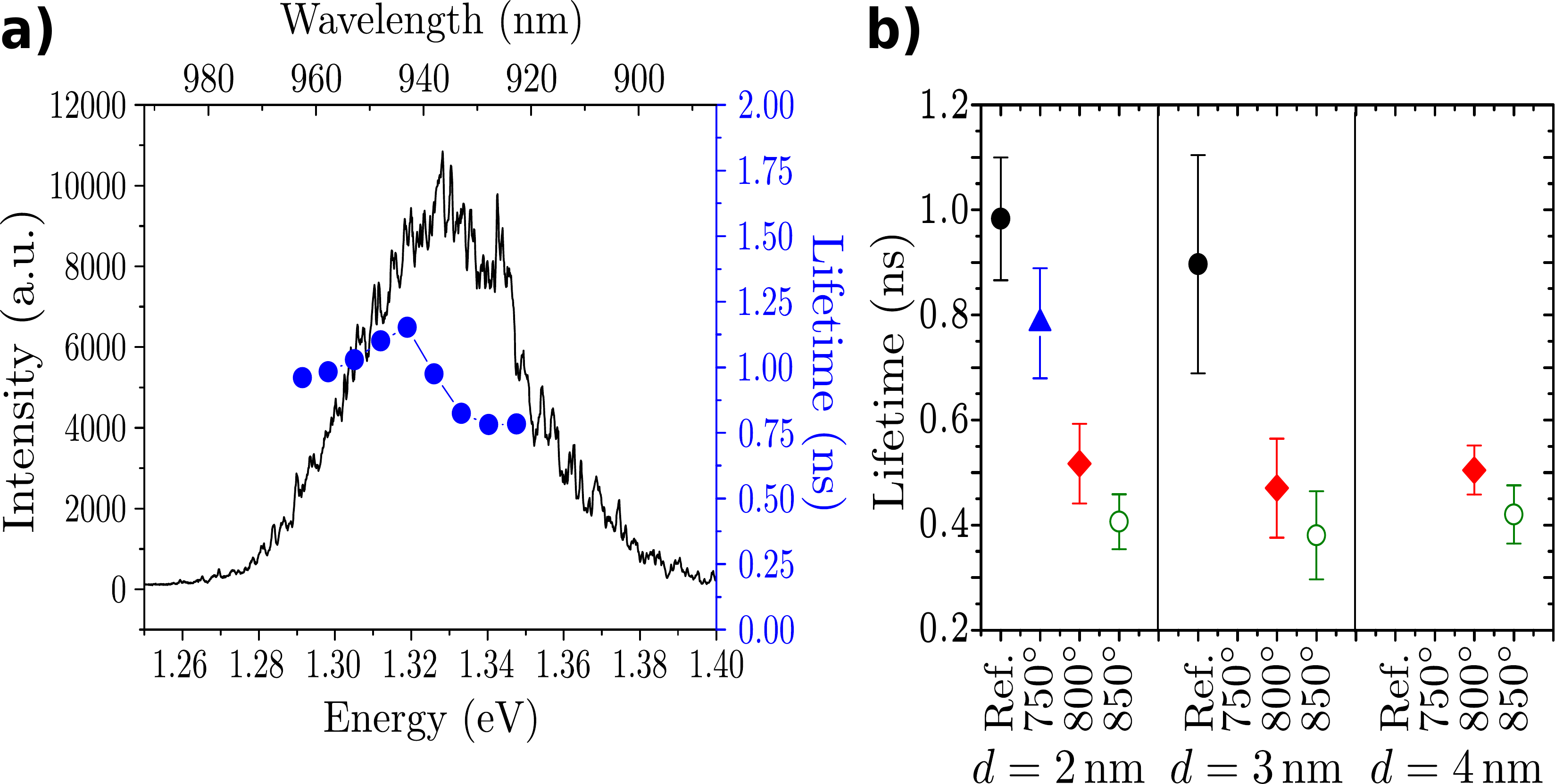}
\caption{a) Exemplaric emission of the QD ensemble of the reference sample with $d= 2~nm$. In addition the lifetimes of the ensemble for different wavelengths are shown. b) Overview of the lifetimes aquired at the spectral positions used for determination of the diamagnetic shift and the Zeeman splitting versus the cap thickness $d$ and the RTA temperature. Higher annealing temperatures lead to a decrease of the mean lifetime of the ensemble and thereby to an increase of the OS.}
\label{fig:lebensdauer}
\end{center}
\end{figure}

According to equation \ref{eq:Oszillatorst�rke_lang} the OS is directly related to the radiative decay time of the quantum emitter. Therefore, measuring the lifetime $\tau$ of the QDs is another way to study the influence of ex-situ RTA on the OS. Fig \ref{fig:lebensdauer} a) exemplarily shows the QD ensemble emission of the reference sample with $d=2~nm$. In this range we performed lifetime measurements by measuring gradually the monochromator filtered emission wavelength (spectral range 0.2~nm) over the whole QD emission spectrum. The lifetimes aquired at the spectral positions used for determination of the diamagnetic shift and the Zeeman splitting are plotted in fig. \ref{fig:lebensdauer} b). 

One cleary sees that a higher annealing temperature results in an enhancement of the spontaneous emission decay rate. Since $f$ is proportional to $\Gamma = 1/\tau$ we note that the annealing procedure is a viable tool to increase the QD's oscillator strenght, as we detail in the next section.

\subsection{Oscillator strength}
\label{os}

\begin{figure}[t]
\begin{center}
\includegraphics[width=0.48\textwidth]{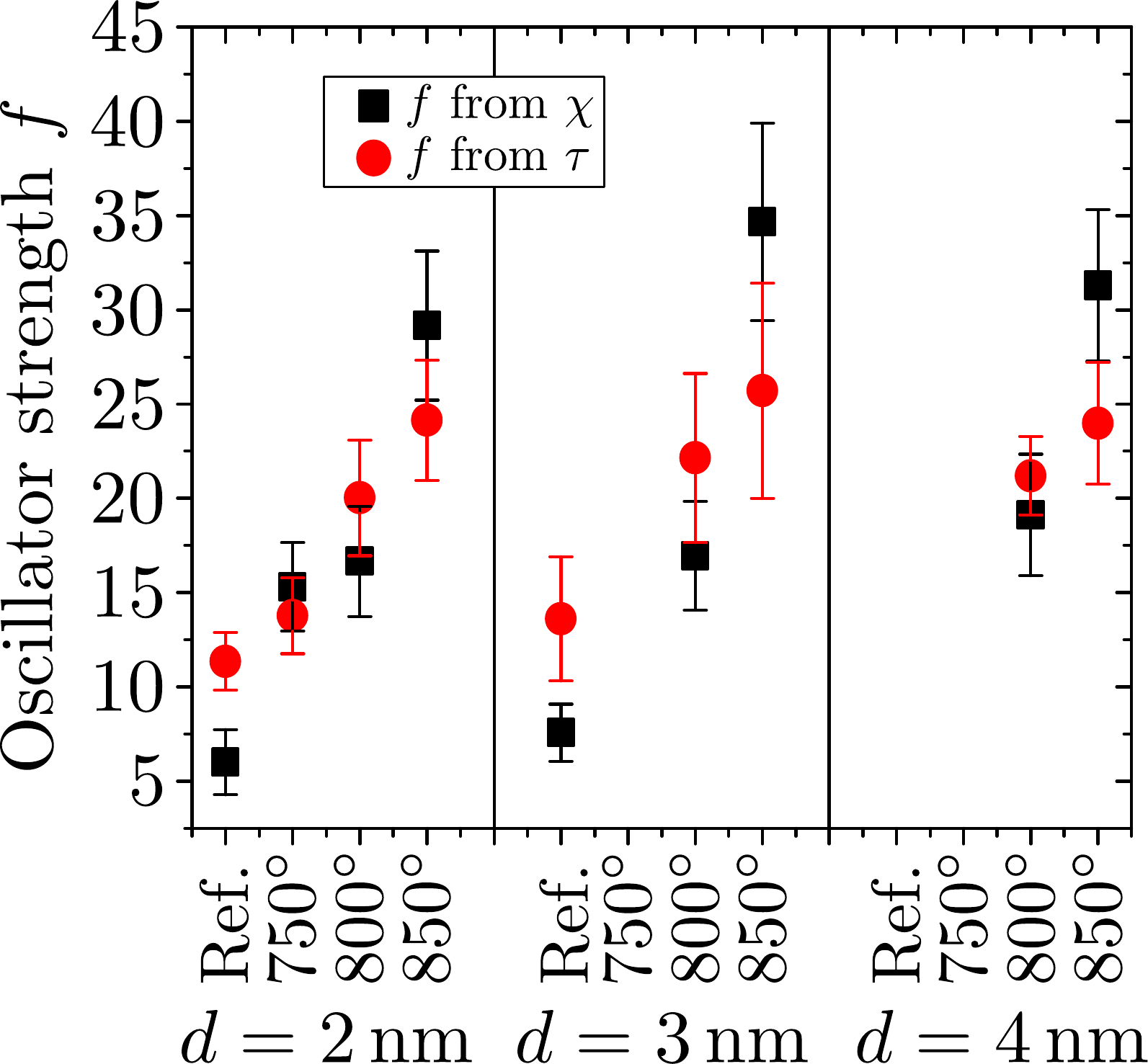}
\caption{Overview of the oscillator strength determined by the lifetime measurements using equation \ref{eq:Oszillatorst�rke_lang} (red, circles) and by the diamagnetic shift $\chi$ using equation \ref{eq:OS_dia} (black, squares), respectively. Both methods show a significant increase of the oscillator strength for higher annealing temperatures and almost all values for $f$ agree within the error bars. The maximum OS $f=25.7\pm 5.7$ from the lifetime measurements as well as the maximum OS $f=34.7\pm 5.2$ from the diamagnetic shift have been obtained for a cap thickness of $d=3~nm$ and an annealing temperature of 850$^{\circ}$~C.}
\label{fig:Oszillatorstarke}
\end{center}
\end{figure}

\begin{table*}[btp]%
\footnotesize
\begin{center}
\begin{tabular}{|c|c|c||c|c|c|c||c|c|}
\hline 
 $d$& Annealing temperature & In content & $|g_X|$ & $\chi \left( \frac{\mu eV}{T^2}\right) $ & \rule[0pt]{0pt}{17pt} $L \approx 2\sqrt{ \left<\rho^2_{e,h}\right>}$\,(nm) & $f$ (from $\chi$)  &$\tau(ns)$& $f$ (from $\tau$) \tabularnewline
\hline 
\hline 
\multirow{4}{*}{2~nm} & not annealed & $45.0\pm5.0$ & $1.71\pm 0.24$ & $6.0\pm1.5$ & $6.8\pm0.8$ & $6.0\pm1.7$ & $0.98\pm0.12$ & $11.4\pm1.5$\tabularnewline
\cline{2-9} 
 & 750$^{\circ}$ C & $36.0\pm5.0$ & $1.14\pm 0.11$ & $12.0\pm0.7$ & $10.0\pm0.4$ & $15.3\pm2.3$ & $0.78\pm0.10$ & $13.8\pm2.0$\tabularnewline
\cline{2-9} 
 & 800$^{\circ}$ C & $34.0\pm5.0$ & $0.86\pm 0.17$ & $12.6\pm1.4$ & $10.3\pm0.6$ & $16.7\pm2.9$ & $0.52\pm0.08$ & $20.0\pm3.1$\tabularnewline
\cline{2-9} 
 & 850$^{\circ}$ C & $28.0\pm5.0$ & $1.24\pm 0.07$ & $19.3\pm0.8$ & $13.0\pm0.3$ & $29.2\pm4.0$ & $0.41\pm0.05$ & $24.1\pm3.2$\tabularnewline
\hline
\hline 
\multirow{3}{*}{3~nm} & not annealed & $51.0\pm5.0$ & $1.62\pm 0.13$ & $8.7\pm1.0$ & $8.0\pm0.5$ & $7.6\pm1.5$ & $0.90\pm0.21$ & $13.6\pm3.3$\tabularnewline
\cline{2-9} 
 & 800$^{\circ}$ C & $39.0\pm5.0$ & $1.48\pm 0.30$ & $14.7\pm1.4$ & $10.9\pm0.6$ & $17.0\pm2.9$ & $0.47\pm0.09$ & $22.1\pm4.5$\tabularnewline
\cline{2-9} 
 & 850$^{\circ}$ C & $22.5\pm5.0$ & $0.92\pm 0.21$ & $20.0\pm1.7$ & $13.5\pm0.6$ & $34.7\pm5.2$ & $0.38\pm0.08$ & $25.7\pm5.7$\tabularnewline
\hline 
\hline
\multirow{2}{*}{4~nm} & 800$^{\circ}$ C & $42.0\pm5.0$ & $0.153\pm 0.038$ & $17.8\pm1.5$ & $11.9\pm0.6$ & $19.1\pm3.2$ & $0.50\pm0.05$ & $21.2\pm2.1$\tabularnewline
\cline{2-9} 
 & 850$^{\circ}$ C & $26.0\pm5.0$ & $0.059\pm 0.034$ & $19.4\pm0.3$ & $13.2\pm0.3$ & $31.3\pm4.0$ & $0.42\pm0.06$ & $24.0\pm3.2$\tabularnewline
\cline{2-9} 
\hline 
\end{tabular}
\caption{Overview over the measured effective Land\'e factor $g_X$, the diamagnetic coefficients $\chi$ and lifetimes $\tau$ as well as the consequent calculated OS $f$. Overall for all cap thicknesses $d$ an increase of the OS for higher RTA temperatures is observable. Taking a reference value of $f=10$ for common not annealed InAs QD samples\cite{Warburton1997} into account the ex-situ RTA process leads to a significant increase beyond a factor of $2$ of the OS.}
\label{tab:os}
\end{center}
\end{table*}

We now give a more detailed analysis of the data. Based on equation \ref{eq:OS_dia} and \ref{eq:Oszillatorst�rke_lang} we can calculate the OS from the experimental data. This calculation requires knowledge of the In content in the QDs. The In content was estimated in the following way: First, Nextnano simulations were carried out for the as-grown QDs, by adjusting the In content of the simulated QD in a way that the calculated transition energies matched the experimentally measured PL energies. The geometric parameters of the QDs were taken from the TEM analysis. The STEM analysis shows that the actual QD height (2.9~nm) is in very good agreement with the nominal height of the capping layer (3~nm). Therefore, 2~nm, 2.9~nm and 4~nm were used as QD heights for the respective samples. A QD width of 26~nm was used for all three nominal capping layer heights since the QD width is not expected to vary significantly between different capping procedures. In a second step we performed a boxcar type smoothing algorithm based on Fick's law to simulate the reduction of In content during the annealing process. The results of this calculations are compiled in the third column of table \ref{tab:os}. Finally, we use these estimations of the In content to calculate the oscillator strength for the acquired data. All other material dependent parameters were taken from literature\cite{Coldren1995,Stier1999,Vurgaftman2001}.

Almost all values for $f$ derived from the diamagnetic shift as well from the lifetime measurements agree within the error bars (see fig. \ref{fig:Oszillatorstarke}). The highest OS $f=34.7\pm 5.2$ derived from the diamagnetic shift has been obtained for a cap thickness of $d=3~nm$ and an annealing temperature of 850$^{\circ}$~C. The maximum OS $f=25.7\pm 5.7$ from the lifetime measurements was calculated also for $d=3~nm$ and a temperature of 850$^{\circ}$~C. For not annealed samples Warbuton et al.\cite{Warburton1997} indicates a OS about 10 which, indeed, is good agreement with our reference samples. %In literature one finds for the diamagnetic coefficient of InAs/GaAs and In$_{0.6}$Ga$_{0.4}$As QDs values of $\chi_{\text{InAs/GaAs}}=10\pm 3~\frac{\mu eV}{T^2}$ \cite{Paskov2000} and $\chi_{\text{In$_{0.6}$Ga$_{0.4}$As}}=8~\frac{\mu eV}{T^2}$ \cite{Reitzenstein2009}, respectively. These values are also comparable to our reference samples. The measured diamagnetic coefficients for annealing temperatures of 800$^{\circ}$~C correspond to those of In$_{0.45}$Ga$_{0.55}$As QDs ($\chi_{\text{In$_{0.45}$Ga$_{0.55}$As}}=20~\frac{\mu eV}{T^2}$)\cite{Reitzenstein2009} that exhibit an OS of about $f=20$. 
Furthermore an increase of the lateral expansion of the QD wave function from $L_{ref}=6.8\pm 0.8~nm$ of the reference sample with $d=2~nm$ to $L_{850^{\circ}}=13.5\pm 0.6~nm$ for $d=3~nm$ and a temperature of 850$^\circ$~C can be extracted from table \ref{tab:os}, where we provide a full list of extracted parameters. In summary we have shown that the mean lateral expansion of the wavefunction increases by about 70\% and the OS has more than doubled due to an ex-situ RTA process.

\section{Summary}
In conclusion with have systematically investigated the influence of an ex-situ rapid thermal annealing process on InAs/GaAs PCA QDs with various cap thicknesses $d$. We studied the emission properties of these QDs in a magnetic field and extracted the effective Land\'e g-factor and the diamagnetic coefficient $\chi$. Both show a strong dependency on the annealing temperature as well as on the cap thickness. From these measurements we could derive the mean lateral expansion of the wavefunction and the oscillator strength $f$ of the QDs. Furthermore we compared the results to time resolved measurements to extract  the oscillator strength $f$ on the QD ensemble. In these studies, we could confirm that RTA has a consistently positive effect on the oscillator strength of an InAs quantum emitter, which makes this technique highly appealing for post growth engineering dynamic emitter properties. We furthermore emphasize that both approaches, namely time resolved spectroscopy and magneto optical measurements yield values which are in very good agreement. 

\section*{Acknowledgment}
The authors acknowledge financial support by the State of Bavaria and the German Ministry of Education and Research (BMBF) within the project Q.com. We thank M. Lermer for sample fabrication.
\appendix
\section{Modeling the effect of annealing}
To simulate the effect of annealing we turned to a boxcar type smoothing algorithm. We started with a quantum dot having uniform In-distribution inside the boundaries of a capped pyramid. Subsequently, this proto-QD is being annealed applying $N$-times the operation
\beq
\begin{array}{ccl}
c_{i,j,k}&\longrightarrow&\frac{1}{7} (c_{i+1,j,k}+c_{i-1,j,k}+c_{i,j+1,k}+\\&&c_{i,j-1,k}+c_{i,j,k+1}+c_{i,j,k-1}+c_{i,j,k})~,
\end{array}
\label{eq:Annealing}
\eeq
with $c$ being the Indium concentration at coordinates $(i,j,k)$ of the numerical grid. The number $N$ refers to the number of smoothing steps. Then, for each $N$ a complete electronic structure calculation is carried out, thus accounting for the corresponding material redistribution, the resulting strain and piezoelectricity. The single particle states are calculated using a 3D implementation of eight-band-\textbf{\textit{k}$\cdot$\textbf{\textit p}} theory and the electron-hole pair lifetime is obtained within the dipole approximation.

\section{Effect of annealing on the optical properties of the samples}

\begin{figure}[btp]
\begin{center}
\vspace{0.5cm}
\includegraphics[width=0.48\textwidth]{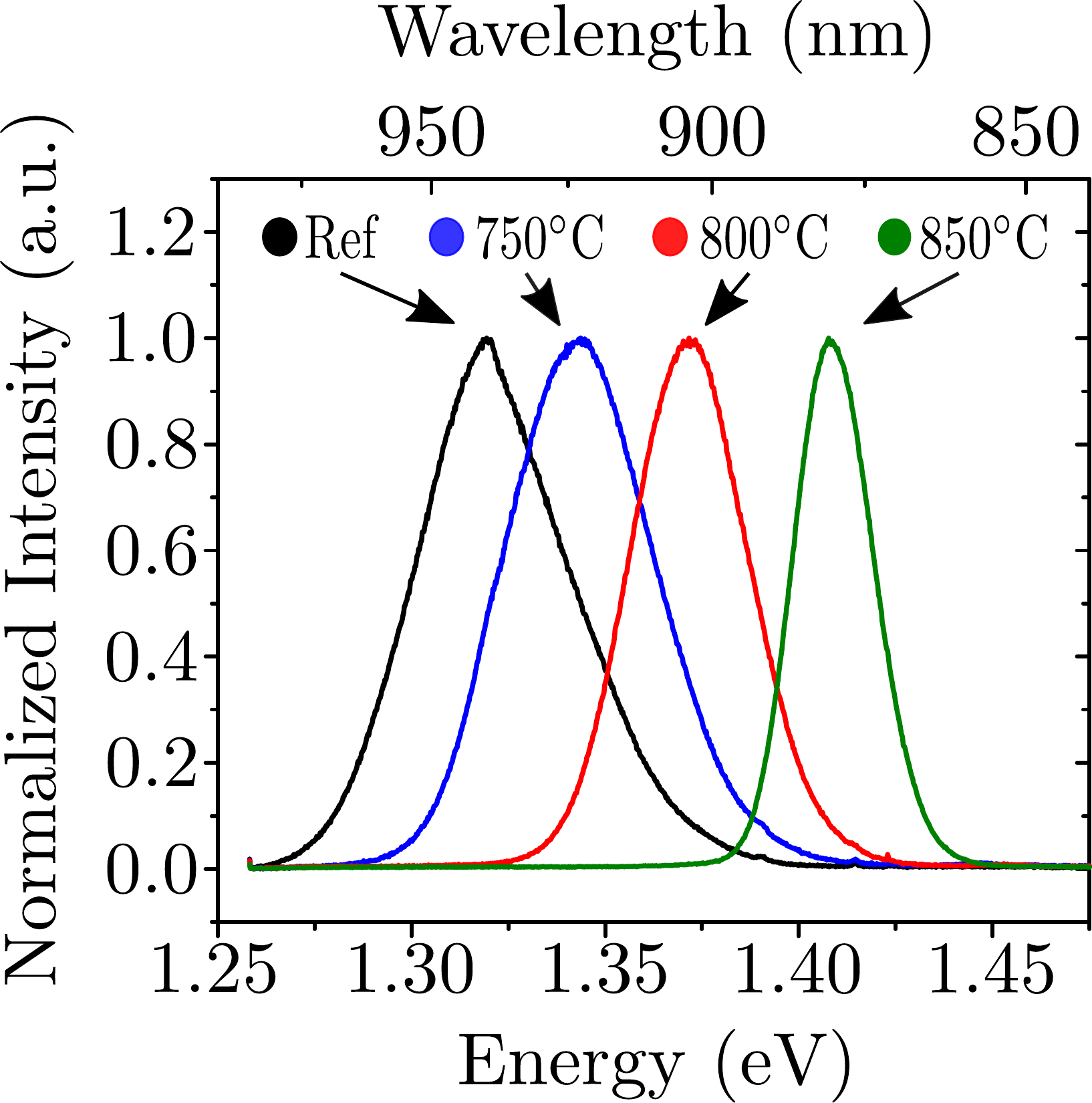}
\caption{Ensemble photoluminescence spectra recorded at 10 K for the sample with a 2 nm high partial cap, which is discussed in the main text. The spectra correspond to the four selected annealing conditions. }
\label{fig:Spektren}
\end{center}
\end{figure}

Ensemble photoluminescence spectra (excitation spot diameter $\approx$ 100 $\mu$m) recorded at 10 K for the sample with a 2 nm high partial cap are plotted in Fig. \ref{fig:Spektren}. The spectra correspond to the four selected annealing conditions, which are also addressed in the main text. From these spectra, the intermixing induced emission blueshift and the narrowing in the ensembles' inhomogeneous broadening becomes obvious, in agreement with previous reports.\cite{Kosogov1996,Fafard1999,Nishi1999,Hsu2000, leon1996effects}.

\begin{table*}[btp]%
\footnotesize
\begin{center}
\begin{tabular}{|c|c|c|c|}
\hline 
 Annealing temperature & $E_{Cent}$\,(eV) & $\Delta_{E}$\,(meV) & FWHM\,(meV) \tabularnewline
\hline 
\hline 
not annealed & 1.321 & --- & 47.9 \tabularnewline
\hline
750$^{\circ}$ C & 1.343 & 22.3 & 45.0 \tabularnewline
\hline
800$^{\circ}$ C & 1.372 & 50.7 & 35.7 \tabularnewline
\hline
850$^{\circ}$ C & 1.409 & 87.9 & 24.1 \tabularnewline
\hline
\end{tabular}
\caption{Peak energy, energy shift and full width at half maximum of the ensemble luminescence extracted from the luminescence spectra plotted in Fig. \ref{fig:Spektren}. }
\label{tab:op}
\end{center}
\end{table*}

\bibliographystyle{apsrev4-1}
\bibliography{Braun}

%\end{document}

%\clearpage

%\noindent Figure Captions

%\begin{figure}[h]
%\centering\includegraphics[width=\textwidth]{Figures/Figure1.eps}
%\caption{\label{fig1} }
%\end{figure}

%\begin{figure}[h]
%\centering\includegraphics[width=\textwidth]{Figures/Figure2.eps}
%\caption{\label{fig2} }
%\end{figure}

%\clearpage

%\begin{figure}[h]
%\centering\includegraphics[width=\textwidth]{Figures/Figure3.eps}
%\caption{\label{fig3} }
%\end{figure}

%\begin{figure}[h]
%\centering\includegraphics[width=\textwidth]{Figures/Figure4.eps}
%\caption{\label{fig4}}
%\end{figure}

%\clearpage

%\begin{figure}[h]
%\centering\includegraphics[width=\textwidth]{Figures/Figure5.eps}
%\caption{\label{fig5}}
%\end{figure}

%\clearpage

%\begin{figure}[h]
%\centering\includegraphics[width=12 cm]{Figures-2.eps}
%\end{figure}
%\centering Unsleber et al. Fig. 2.

%\clearpage

%\begin{figure}[h]
%\centering\includegraphics[width=12 cm]{Figures-3.eps}
%\end{figure}
%\centering Unsleber et al. Fig. 3.

%\clearpage

%\begin{figure}[h]
%\centering\includegraphics[width=12 cm]{Figures-4.eps}
%\end{figure}
%\centering Unsleber et al. Fig. 4.

\end{document}